\documentclass[
    aps,
    pra,
    reprint,
    amsmath,
    amssymb,
    superscriptaddress
]{revtex4-2}

\usepackage{ragged2e}
\usepackage{graphicx}
\usepackage{dcolumn}
\usepackage{bm}
\usepackage{orcidlink}
\usepackage{braket}

\begin{document}

\preprint{APS/123-QED}

\title{\textbf{Implementation of a 1D Gaussian-modulated CV-QKD under finite-size parameter estimation} 
}%

\author{Christiano M. S. Nascimento\,\orcidlink{0009-0000-2160-8742}}
\email{christiano.moreira@fieb.org.br}
\affiliation{QuIIN -- Quantum Industrial Innovation, EMBRAPII CIMATEC Competence Center in Quantum Technologies, SENAI CIMATEC, Av. Orlando Gomes 1845, Salvador, 41650-010, BA, Brazil}
\affiliation{NITeQ, Department of Electrical Engineering, Pontifical Catholic University of Rio de Janeiro, 22451-900 Rio de Janeiro, RJ, Brazil}

\author{Maron F. Anka\,\orcidlink{0000-0001-6720-4620}}
\affiliation{QuIIN -- Quantum Industrial Innovation, EMBRAPII CIMATEC Competence Center in Quantum Technologies, SENAI CIMATEC, Av. Orlando Gomes 1845, Salvador, 41650-010, BA, Brazil}

\author{Gabriel A. V. Menezes}
\affiliation{QuIIN -- Quantum Industrial Innovation, EMBRAPII CIMATEC Competence Center in Quantum Technologies, SENAI CIMATEC, Av. Orlando Gomes 1845, Salvador, 41650-010, BA, Brazil}

\author{Davi J. G. de Sousa\,\orcidlink{0009-0009-1946-5573}}
\affiliation{QuIIN -- Quantum Industrial Innovation, EMBRAPII CIMATEC Competence Center in Quantum Technologies, SENAI CIMATEC, Av. Orlando Gomes 1845, Salvador, 41650-010, BA, Brazil}

\author{Nelson A. Neto\,\orcidlink{0000-0003-2278-1082}}
\affiliation{QuIIN -- Quantum Industrial Innovation, EMBRAPII CIMATEC Competence Center in Quantum Technologies, SENAI CIMATEC, Av. Orlando Gomes 1845, Salvador, 41650-010, BA, Brazil}

\author{Artur A. Matoso\,\orcidlink{0000-0003-4039-0544}}
\affiliation{QuIIN -- Quantum Industrial Innovation, EMBRAPII CIMATEC Competence Center in Quantum Technologies, SENAI CIMATEC, Av. Orlando Gomes 1845, Salvador, 41650-010, BA, Brazil}

\author{Gustavo C. Amaral\,\orcidlink{0000-0001-6071-7458}}
\affiliation{Quantum Technology Department, TNO -- The Netherlands Organization for Applied Scientific Research, 2628CK Delft, The Netherlands}

\author{Valéria L. Silva\,\orcidlink{0000-0001-5466-7933}}
\affiliation{QuIIN -- Quantum Industrial Innovation, EMBRAPII CIMATEC Competence Center in Quantum Technologies, SENAI CIMATEC, Av. Orlando Gomes 1845, Salvador, 41650-010, BA, Brazil}

\author{Guilherme P. Temporão\,\orcidlink{0000-0002-5244-3394}}
\affiliation{NITeQ, Department of Electrical Engineering, Pontifical Catholic University of Rio de Janeiro, 22451-900 Rio de Janeiro, RJ, Brazil}

\date{\today}

\begin{abstract}
Continuous-variable quantum key distribution (CV-QKD) is a promising technology for quantum-secure communications, offering high secret-key rates and straightforward integration with coherent optical communication systems. Conventional Gaussian-modulated CV-QKD encodes information in both optical-field quadratures, increasing requirements for state preparation, modulation, and calibration, \textcolor{black}{while} potentially introducing additional preparation imperfections. Unidimensional (1D) Gaussian modulation offers a simpler alternative by encoding information in only one quadrature, while the conjugate quadrature constrains the protocol's security. However, an integrated experimental 1D implementation combining independent-laser coherent reception with practical reconciliation, finite-size parameter estimation, and privacy amplification remains challenging. Here, we experimentally demonstrate a 1D Gaussian-modulated CV-QKD system over 12~km of standard single-mode fiber using a \textcolor{black}{local local oscillator} (LLO) architecture with independent lasers and an integrated offline classical post-processing chain. The implementation includes coherent detection, digital signal processing, information reconciliation, parameter estimation, physicality-constrained security analysis, and privacy amplification. The 1D system achieves a privacy-amplified rate of $1.421$~Mbit/s under the asymptotic security treatment and \textcolor{black}{$0.364$~Mbit/s under the finite-size scenario.} These results \textcolor{black}{establish an experimental baseline for} the performance--complexity trade-off associated with unidimensional Gaussian modulation and provide an experimental basis for developing lower-complexity CV-QKD transmitters.
\end{abstract}

\maketitle

\section{Introduction}
\label{sec:introduction}

Quantum key distribution (QKD) enables two distant parties, Alice and Bob, to generate secret keys whose security rests on the laws of quantum physics rather than assumptions about computational complexity \cite{Gisin:2002,Thew:2007,bennett2020quantum,wolf2021quantum,anka2026introductory}. Among the different approaches to QKD, continuous-variable quantum key distribution (CV-QKD) has attracted considerable attention because it can leverage coherent optical communication technologies, offering high secret-key rates over metropolitan distances and facilitating integration with existing optical communication infrastructure \cite{grosshans2001continuous,weedbrook2012gaussian,juvencio2026towards,hajomer2025coexistence,melgar2024coexistence}. In conventional Gaussian-modulated (GM) CV-QKD, hereafter referred to as two-dimensional (2D) Gaussian modulation, Alice independently modulates both optical-field quadratures. Bob then measures the received states using homodyne or heterodyne detection \cite{laudenbach2018continuous,weedbrook2004quantum,jouguet2013experimental}.

The unidimensional (1D) GM CV-QKD protocol proposed by Usenko and Grosshans \cite{usenko2015unidimensional} provides an alternative in which information is encoded in only one optical-field quadrature. The conjugate quadrature remains nominally unmodulated and does not directly contribute to the raw key. Still, its measured statistics constrain the physically admissible Alice--Bob covariance matrices and therefore the information potentially available to Eve. By reducing the dimensionality of state preparation, the 1D protocol offers the prospect of lower transmitter complexity, at the cost of a more constrained security analysis and increased sensitivity to finite-size statistical uncertainty and experimental imperfections.

Several experimental implementations of 1D and related single-quadrature CV-QKD protocols have been reported \cite{gehring2015single,wang2017experimental,shen2019free,nandan2026experimental,guo2026pilot}. However, to the best of our knowledge, no experimental 1D GM CV-QKD implementation has combined independent-laser coherent reception, practical information reconciliation, finite-size parameter estimation (PE), physicality-constrained security analysis, privacy amplification, and an integrated classical post-processing chain. Here, we implement such system in an optical fiber of 12~km with independent lasers and RF heterodyne detection \textcolor{black}{and compare the measured key rates, under asymptotic and finite-size frameworks.}

\begin{figure*}[t]
    \centering
    \includegraphics[width=\textwidth]{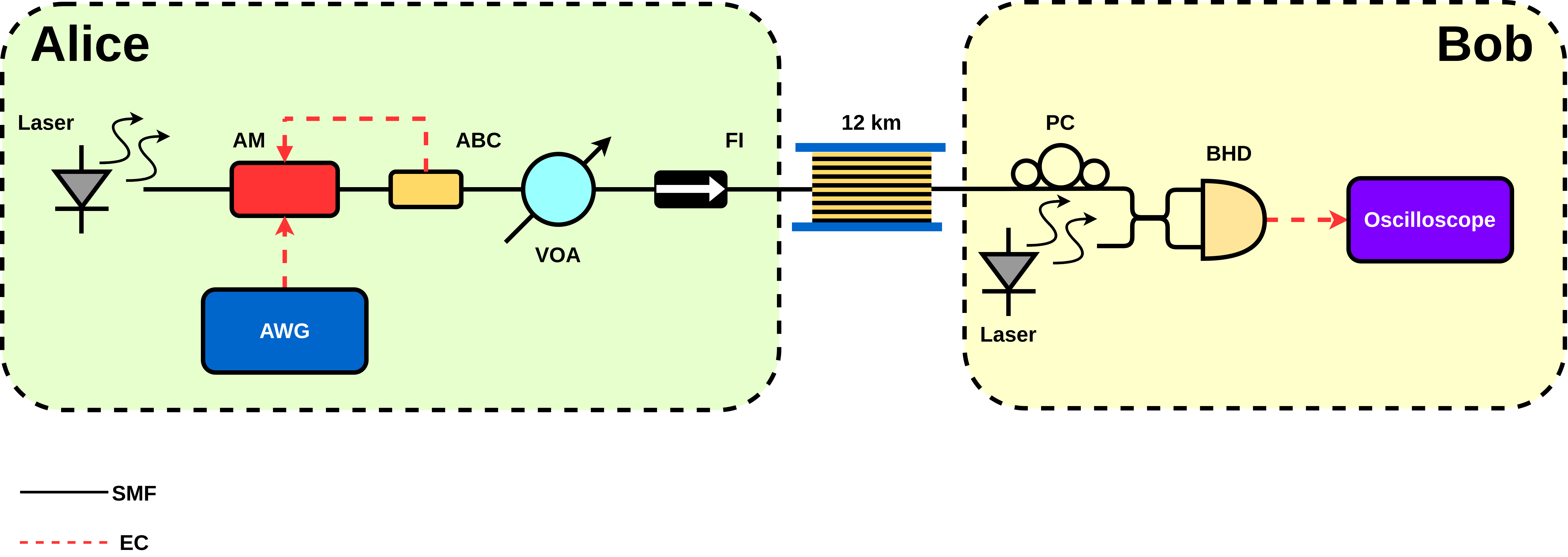}
    \caption{\justifying Experimental setup of LLO CV-QKD. Alice’s station consists of a continuous-wave (CW) laser operating at 1550 nm and an in-phase and quadrature modulator (IQM) with an automatic bias controller (ABC) driven by an arbitrary waveform generator (AWG) with 16-bit resolution and a sampling rate of 3.125 GSa/s. A variable optical attenuator (VOA) is placed after the amplitude modulator (AM) modulator (at the exit of Alice's station) to adjust the optical power and enforce the few-photon regime. A 12 km single-mode fiber (SMF) constitutes the quantum channel. At Bob’s station, a polarization controller (PC) allows one to align the polarization of the incoming signal to the LO signal generated from an independent CW laser; both overlap at a balanced beam splitter. The differential signal was detected and digitized using a balanced homodyne detector (BHD), followed by a 12-bit oscilloscope with a matched sampling rate of 3.125 GSa/s. EC: electrical connection.}
    \label{fig:experimental_setup}
\end{figure*}

\section{Methodology}
\label{Method}

In conventional GM CV-QKD, both optical-field quadratures are modulated, allowing direct estimation of the corresponding Alice--Bob correlations. In contrast, the 1D protocol modulates only one quadrature, here denoted by $X$, while the conjugate quadrature $P$ remains unmodulated \cite{usenko2015unidimensional}. Consequently, the Alice--Bob correlation $C_P$ associated with the unmodulated quadrature cannot be directly estimated from Alice's modulation data and must instead be constrained by the physicality condition of the covariance matrix.

For each physically admissible covariance matrix, we evaluate Eve's \textcolor{black}{information, i.e., the Holevo information ($\chi_{BE}$),} from its symplectic spectrum under the Gaussian collective-attack model \cite{weedbrook2012gaussian,usenko2015unidimensional}. \textcolor{black}{In the 1D protocol only the X-quadrature outcomes $x_B$ form the raw key, whereas the P-quadrature outcomes are used solely to estimate Bob's unmodulated-quadrature variance and are never disclosed. Eve's information is therefore bounded by $\chi(x_B:E)$, for which Bob's key measurement is the X marginal of the RF-heretordyne detection, i.e., a homodyne measurement with efficiency $\eta_B/2$ and trusted electronic noise \cite{jouguet2013experimental,gehring2015single} (see Appendix). If both heterodyne outcomes entered the key, or if $p_B$ were disclosed, Eve's information would have to be bounded by the larger quantity $\chi(x_B,p_B:E)$ \cite{usenko2025shared}.} In the asymptotic regime, we maximize $\chi_{BE}$ over the physically allowed values of $C_P$. In the finite-size regime, this optimization additionally incorporates conservative confidence intervals for the experimentally estimated channel transmittance, excess noise, and unmodulated-quadrature variance, accounting for statistical uncertainties in the security analysis, \textcolor{black}{and the secret key fraction is further reduced by privacy amplification correction $\Delta (n)$ \cite{leverrier2010finite}.} The Appendix provides a detailed formulation of the 1D covariance matrix, the physicality constraints, \textcolor{black}{the Holevo bound,} and the finite-size procedure.

\begin{figure*}[t]
    \centering
    \includegraphics[width=\textwidth]{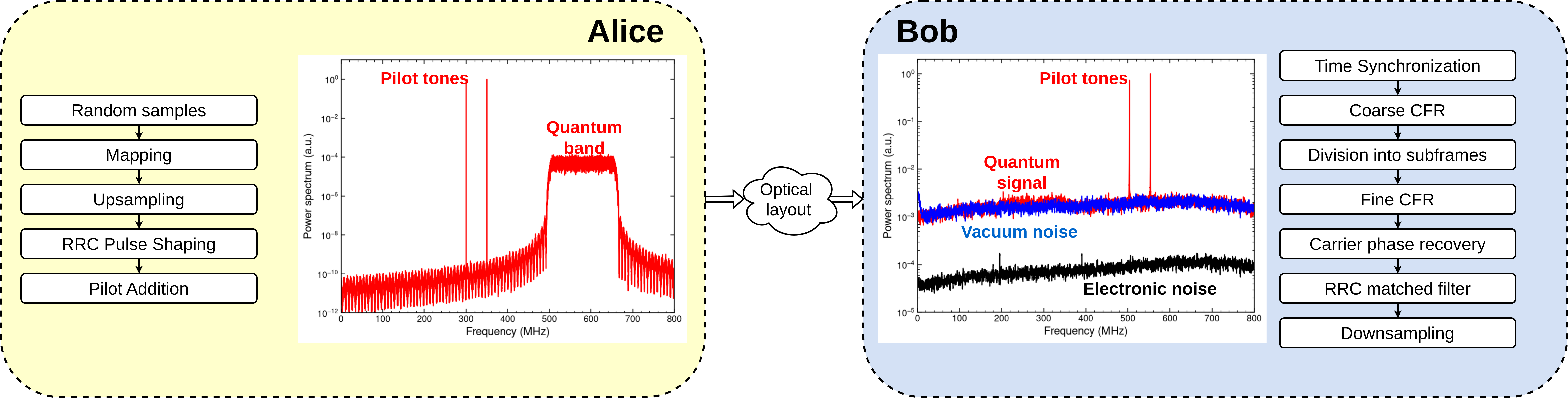}
    \caption{\justifying DSP chain used in the experiment, including transmitter-side Gaussian symbol generation and pulse shaping, pilot-tone and CAZAC insertion, and receiver-side synchronization, carrier recovery, matched filtering, and downsampling. CAZAC: constant amplitude zero autocorrelation; CFR: carrier-frequency recovery; RRC: root-raised cosine.}
    \label{fig:dsp_chain}
\end{figure*}

Figure~\ref{fig:experimental_setup} schematically illustrates the 1D CV-QKD architecture using an amplitude modulator (AM). Alice and Bob use independent narrow-linewidth ($\sim1$~kHz) CW lasers centered at 1550~nm in an LLO architecture. Although the 1D protocol requires modulating only one optical-field quadrature, the experimental implementation used an IQ modulator driven by a two-channel 16-bit AWG operating at 3.125~GSa/s. The IQ modulator was operated in single-sideband (SSB) mode to compensate for the optical-frequency separation between the independent lasers, shifting the signal toward a lower intermediate frequency within the detector bandwidth. Therefore, the IQ modulator was used because of experimental constraints rather than an intrinsic requirement of the 1D protocol. An automatic bias controller stabilizes the modulator, while a variable optical attenuator (VOA) sets the transmitted modulation level. A Faraday isolator reduces back-reflections and susceptibility to Trojan-horse attacks \cite{do2025side}. The Gaussian modulation sequence is generated pseudorandomly on a standard PC and is used here as a proof-of-concept source; a practical transmitter can replace it with a QRNG without changing the optical architecture.

The quantum channel consists of 12~km of SMF with an average attenuation of 0.185~dB/km, corresponding to a total loss of 2.22~dB. Bob performs RF-heterodyne detection by interfering the received signal with an independent LLO whose optical-frequency offset from Alice is approximately 850~MHz \cite{hajomer2024long,pietri2024qosst}. Because this frequency separation approaches the nominal 1~GHz bandwidth of the balanced detector, the SSB configuration lets the quantum signal be detected at a lower intermediate frequency, avoiding operation near the detector's bandwidth limit. The interference signal is detected using a 1~GHz balanced detector and digitized at 3.125~GSa/s for offline processing.

Before the QKD acquisitions, we independently calibrated the modulation level in a back-to-back configuration with the 12~km fiber removed. These measurements provided the mean photon number $n_A$ and modulation variance $V_M$, which were then fixed for channel parameter estimation. We obtained Bob's shot-noise and electronic-noise references from dedicated vacuum-plus-electronic and electronic-only acquisitions, respectively. We obtained the shot-noise contribution from the difference between the corresponding variances. The detection efficiency was $\eta_B=0.80$. Throughout the security analysis, we treat the calibrated detector efficiency and electronic noise according to a trusted-receiver model. \textcolor{black}{Both noise references are expressed in units of the shot noise of one heterodyne output, i.e. per quadrature, and the input-reference excess noise is obtained from Bob's estimate as $\xi_{\text{in}} = 2 \xi_B/\eta_B T$ (see Appendix).}

Figure~\ref{fig:dsp_chain} summarizes the DSP. Alice maps the Gaussian symbols, upsamples and RRC pulse-shapes the sequence, and adds two pilot tones and a CAZAC synchronization sequence. Bob performs CAZAC-based timing synchronization, coarse carrier-frequency recovery, subframe division, fine frequency and phase recovery, matched filtering, and symbol-rate downsampling. During phase recovery, a fraction of the quantum symbols is sacrificed to estimate and compensate for phase fluctuations.

Information reconciliation uses reverse eight-dimensional MDR combined with MET-LDPC codes \cite{mani2021multiedge}. LDPC decoding is performed over blocks containing approximately $10^6$ reconciled symbols. We evaluate the reconciliation efficiency $\beta$ from the actual error-correction leakage of successfully decoded blocks, while failed blocks contribute to the FER.

We perform information reconciliation before parameter estimation. We retain successfully decoded and verified blocks for privacy amplification. At the same time, all recovered quantum symbols, including those from failed decoding blocks, contribute to channel characterization following the post-reconciliation strategy of Jain \textit{et al}~\cite{jain2022practical}. The information disclosed for parameter estimation is treated as public and does not contribute to the final privacy-amplified key.

\textcolor{black}{We} construct two-sided confidence intervals for $T_X$ and $V_P^B$, together with an upper confidence bound for $\xi_{B,X}$. Each confidence bound is assigned a failure probability not exceeding $\delta_{\mathrm{PE}}=10^{-10}$, resulting in an overall parameter-estimation failure probability bounded by $3\delta_{\mathrm{PE}}$ through the union bound \cite{leverrier2010finite,wang2017experimental}. We use the resulting confidence region in the physicality-constrained optimization described in the Appendix. \textcolor{black}{The finite-size regime additionally includes the privacy amplification correction}
\begin{equation}
\textcolor{black}{\Delta (n) = 7 \sqrt{\dfrac{\log_2(2/\overline{\epsilon})}{n}},}
\end{equation}
\textcolor{black}{in the form of Ref.\cite{wang2017finite}, where $n = (1 - \mathrm{FER})N_{\mathrm{PE}}$ is the number of symbols of the succesfully reconciled blocks that enter privacy amplification and $\overline{\epsilon} = 10^{-10}$ is the smoothing parameter.} In this work, the treatment is restricted to finite-size parameter-estimation \textcolor{black}{privacy amplification corrections} under the Gaussian collective-attack model and does not constitute a complete composable finite-key security proof.

Table~\ref{tab:experimental_comparison} summarizes the main experimental parameters and secret-key performance. The dataset consisted of 82 independently processed frames with approximately $10^6$ transmitted states per frame, \textcolor{black}{of which 78,562,560 recovered symbols entered parameter estimation and reconciliation.}

\begin{table}[htpb]
    \caption{\label{tab:experimental_comparison}
    Experimental parameters and performance of the 1D GM CV-QKD implementation.}
    \begin{ruledtabular}
        \begin{tabular}{cc}
            Parameter & Experimental value \\
            \hline
            Physical frames & 82 \\
            $\eta_{\mathrm{B}}$ & 0.8 \\
            $\braket{n_A}$ (mean photons number/symbol) & 0.320 \\
            $V_M$ (SNU) & 1.280 \\
            $T$ & 0.5740 \\
            $\xi_{\mathrm{B}}$ (mSNU) & 7.222 \\
            $\xi_{\mathrm{in}}$ (mSNU) & 31.456 \\
            $v_{\mathrm{el}}$ (mSNU) & 34.619 \\
            $V_{\mathrm{P}}^{\mathrm{B}}$ (SNU) & 1.01283 \\
            $\beta$ & 0.9316 \\
            FER (\%) & 32.62 \\
            $N_{\mathrm{total}}$ & 90,20,000 \\
            $N_{\mathrm{PE}}$ & 78,562,560 \\
            $N_{\mathrm{excluded}}$ & 3,437,440 \\
            $\delta_{\mathrm{PE}}$ & $10^{-10}$ \\
            $\overline{\epsilon}$ & $10^{-10}$ \\
            $\Delta (n)$ (bit/symbol) & $5.63 \times 10^{-3}$ \\
            SKR asymptotic (Mbit/s) & 1.421 \\
            SKR finite-size (Mbit/s) & 0.364 \\
        \end{tabular}
    \end{ruledtabular}

    \vspace{0.1cm}

    \begin{minipage}{\columnwidth}
        \footnotesize
        \justifying
        $N_{\mathrm{excluded}}$ -- Number of symbols excluded
        during DSP;
        $N_{\mathrm{PE}}$ -- Number of symbols used for parameter
        estimation, including those from unsuccessful
        reconciliation blocks.
        $N_{\mathrm{total}}$ -- Transmitted symbols in the block
        ($10^6$ per physical frame, including pilot tonesm
        the CAZAC sequence and DSP exclusions), which define
        $t_{\mathrm{data}}$ in Eq.~\eqref{eq1}.
    \end{minipage}
\end{table}

Finally, successfully reconciled and verified blocks are compressed by universal Toeplitz hashing \cite{laudenbach2018continuous,juvencio2026towards}. The experimental secret key rate \textcolor{black}{(SKR)} is

\begin{equation}
    \mathrm{SKR}_{\mathrm{exp}} = \frac{\ell}{t_{\mathrm{data}}} \text{,}
    \label{eq1}
\end{equation}

\noindent where $\ell$ is the data output length and $t_{\mathrm{data}}$ is the total quantum-signal transmission time associated with the processed acquisitions, including acquisitions that do not yield reconciled key material.

\section{Results}
\label{results}

The calibrated mean photon number was $n_A^{\mathrm{1D}}=0.320$ photons per symbol. Under the adopted normalization, $V_M=4n_A$. The mean input-referred excess noises over the first 80 valid frames were $31.444\pm4.412$~mSNU, with no systematic drift observed over the acquisition sequence.

\begin{figure}[htbp]
    \centering
    \includegraphics[width=\columnwidth]{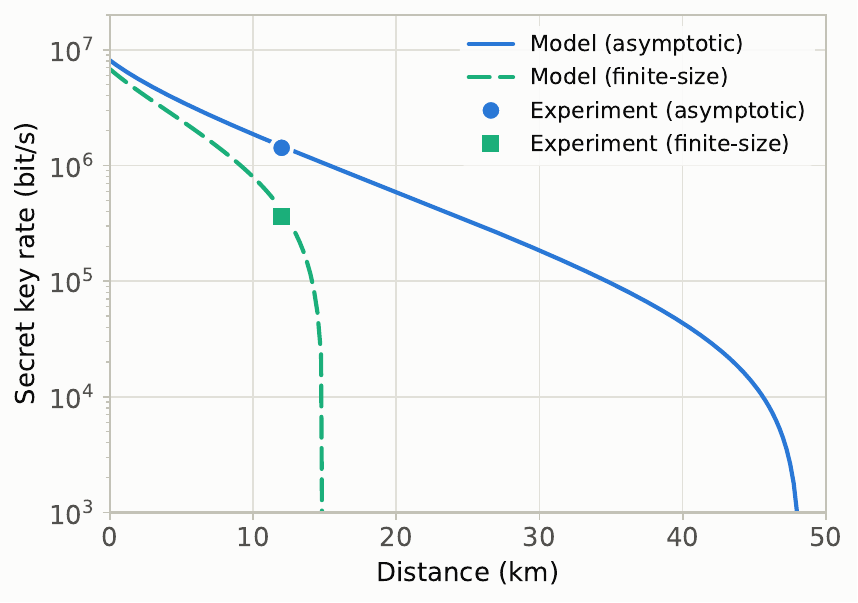}
    \caption{\justifying Secret-key rate versus fiber distance for the 1D implementation. The curves are security-model projections using the experimentally recovered operating parameters and fixed aggregate reconciliation performance, \textcolor{black}{with the block size and confidence levels of the data set ($N_{\mathrm{PE}} = 7.9 \times 10^7$, $\delta_{\mathrm{PE}} = 10^{-10}$, $n = 5.3 \times 10^7$, $\overline{\epsilon} = 10^{-10}$).} \textcolor{black}{Markers indicate the experimental results at 12~km: circle (square) for the asymptotic (finite-size) case.}}
    \label{fig:skr_distance}
\end{figure}

For the analytical performance curves, the finite-size secret-key rate is written as
\begin{equation}
    \mathrm{SKR}_{\mathrm{theory}} = R_s(1 - \mathrm{FER}) \textcolor{black}{\dfrac{N_{\mathrm{PE}}}{N_{\mathrm{total}}}} \left( \beta I_{AB} - \chi_{BE} \right) \textcolor{black}{- \Delta (n)} \text{,}
\end{equation}
\textcolor{black}{where $N_{\mathrm{PE}}/N_{\mathrm{total}}$ accounts for the symbols of each physical frame that do not enter reconciliation, so that the projection is comparable with Eq.~\eqref{eq1}.}

Figure~\ref{fig:skr_distance} shows the resulting SKR as a function of transmission distance. The markers indicate the experimental post-processing results at 12~km, whereas the curves are security-model projections using the experimentally recovered operating parameters. The aggregate experimental $\beta$ and FER are held fixed as distance varies, so the curves should not be interpreted as experimentally validated operating points away from 12~km. \textcolor{black}{ The asymptotic regime reaches 48.3 km, whereas the finite-size case falls to zero at 14.9 km: the correction $\Delta (n)$ of the present block is independent of the channel and removes 0.516~Mbit/s at every distance, while the attenuation effect decrease the SKR quickly with the channel loss.}

\begin{figure}[htpb]
    \centering
    \includegraphics[width=\columnwidth]{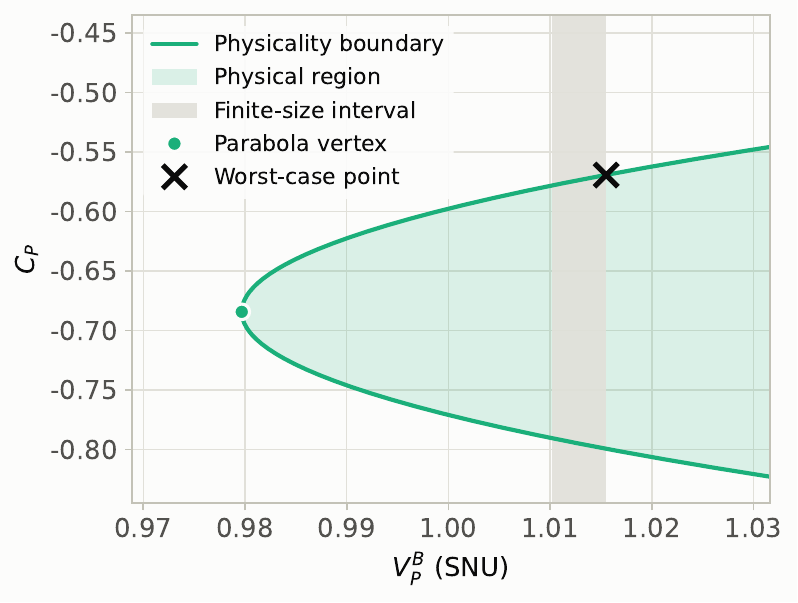}
    \caption{\justifying Physicality-constrained region for the unknown Alice--Bob correlation $C_{\mathrm{P}}$ associated with the unmodulated quadrature, where $V_{\mathrm{P}}^{\mathrm{B}}$ denotes Bob's measured variance of the unmodulated $P$ quadrature. \textcolor{black}{The curve is the physicality boundary of Eq.~(S5) at finite-size wort case $(T_x,\xi_X) = (0.5756, 36.04 mSNU)$.The green shaded area is the physical region, the grey band is the confidence interval of $V_{\mathrm{P}}^{\mathrm{B}}$, and the cross is the worst-case point ($V_{\mathrm{P}}^{\mathrm{B}}$, $C_{\mathrm{P}}$) = (1.0155, -0.5694).} The experimentally recovered $V_{\mathrm{P}}^{\mathrm{B}}$ restricts the set of admissible covariance matrices, and Eve's Holevo information is maximized over this physical region.}
    \label{fig:physicality}
\end{figure}

Figure~\ref{fig:physicality} illustrates the physicality constraint imposed on the unknown correlation $C_P$. The Appendix provides the analytical expressions defining the admissible region and its boundary. In the finite-size treatment, the statistical uncertainty of the experimentally estimated parameters further enlarges the admissible region over which Eve's Holevo information is maximized. \textcolor{black}{The worst case is not found at the lower transmittance bound: because both the excess-noise bound and the physicality region depend on $T_{\mathrm{X}}$, Eve's information is maximized at the upper bound $T_{\mathrm{X}} = 0.5756$, with $\xi_{\mathrm{X}} = 36.04$, $V_{\mathrm{P}}^{\mathrm{B}} = 1.0155$, and $C_{\mathrm{P}} = -0.5694$, where the admissible correlation is largest.}

At 12~km, the implementation achieved $1.421$~Mbit/s asymptotically and \textcolor{black}{$0.364$}~Mbit/s under finite-size regime. The theoretical rates at the same operating point were \textcolor{black}{$1.469$} and \textcolor{black}{0.412}~Mbit/s for the asymptotic and finite-size treatments, respectively. \textcolor{black}{The experimental rates lies 48 kbit/s below both projections. In relative terms, the difference is $3\%$ of the asymptotic rate and $13\%$ of the finite-size rate, while the finite-size approach reduces the experimental rate by 74.4\%.} 

\section{Discussion}
\label{discussion}

\begin{figure}[htpb]
    \centering
    \includegraphics[width=\columnwidth]{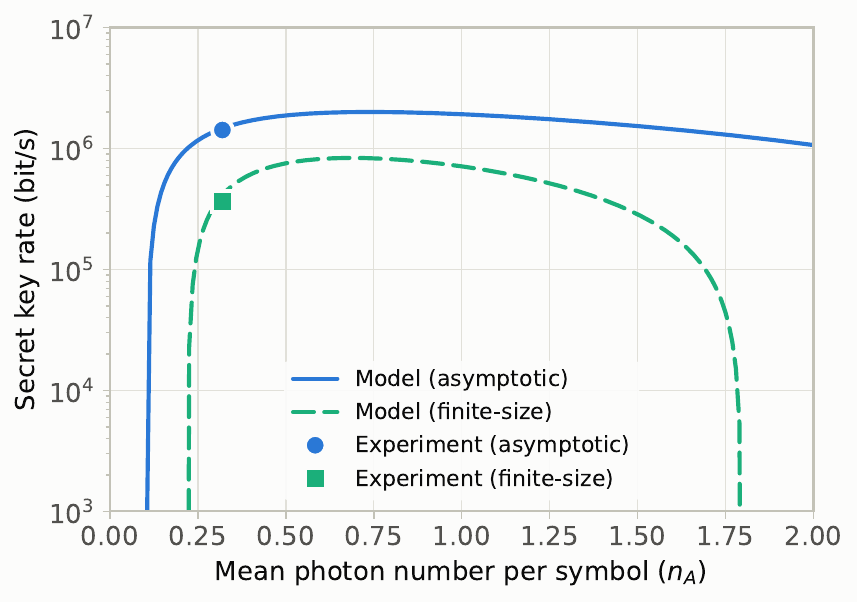}
    \caption{\justifying Asymptotic (solid) and finite-size (dashed) SKR at 12~km as a function of the mean photon number per transmitted symbol $n_A$. Markers indicate the experimental operating point under the two treatments.}
    \label{fig:skr_na}
\end{figure}

To further investigate the performance potential of the 1D protocol, we further investigate how one can improve its capacity. Figure~\ref{fig:skr_na} shows the asymptotic \textcolor{black}{and finite-size} SKR at 12~km as a function of the mean photon number per transmitted symbol, $n_A$, \textcolor{black}{ate the experimental excess-noise level and block size.} Unless otherwise stated, the numerical projections retain the aggregate experimental reconciliation efficiency and FER.

\textcolor{black}{Increasing $n_{\mathrm{A}}$ improves the SKR until it reaches an optimum operating region: the simulation peaks near $n_{\mathrm{A}} \approx 0.75~(V_{\mathrm{M}} \approx 2.0 \text{ SNU})$ asymptotically, at 2.00 Mbit/s, and near $n_{\mathrm{A}} \approx 0.70~(V_{\mathrm{M}} \approx 2.8 \text{ SNU})$ with the finite-size picture, at 0.84 Mbit/s, both above the experimental $n_{\mathrm{A}} = 0.32$.  Beyond the optimum, the finite-size rate falls faster than the asymptotic one and vanished at $n_{\mathrm{A}} \approx 1.8$: the penalty of the confidence region grows with the modulation variance, from $5.9 \times 10^{-3}$ bit/symbol at $n_{\mathrm{A}} = 0.32$ to $8.1 \times 10^{-3}$ bit/symbol at $n_{\mathrm{A}} = 1.8$, because the admissible interval of $C_{\mathrm{B}}$ widens with $V_{\mathrm{M}}$, while $\Delta (n)$ is fixed.} Experimentally, however, $n_A$ and $\xi_{\mathrm{in}}$ are not necessarily independent; larger modulation energies may increase sensitivity to modulator nonlinearities, calibration errors, or residual DSP impairments. The curves should therefore be interpreted as fixed-noise optimization studies rather than as evidence that increasing $n_A$ alone improves experimental performance.

\begin{figure}[htpb]
    \centering
    \includegraphics[width=\columnwidth]{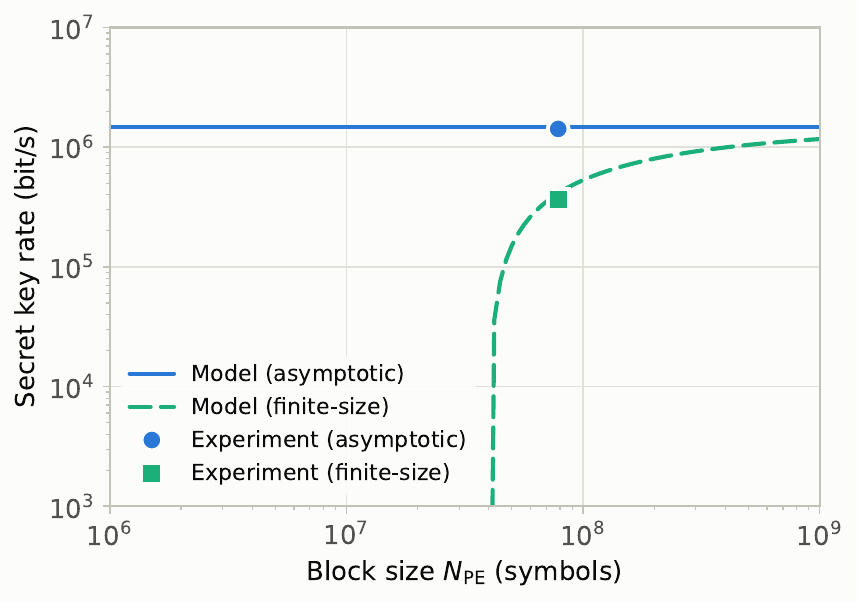}
    \caption{\justifying \textcolor{black}{Asymptotic (solid) and finite-size (dashed) SKR at 12 km as a function of the block size $N_{\mathrm{PE}}$ for the experimental parameters values of Table \ref{tab:experimental_comparison}. Markers indicate the experimental block of $7.9 \times 10^7$ symbols under the two regimes.}}
    \label{fig:skr_block}
\end{figure}

\textcolor{black}{Figure \ref{fig:skr_block} shows the dependence of the rate at 12 km on the block size for the experimental parameters values in Table \ref{tab:experimental_comparison} and for the theoretical simulation under the same assumptions. The asymptotic rate does not depend on the block size. The finite-size key becomes positve only above $N_{\mathrm{PE}} \approx 4.0 \times 10^7$ symbols; at the experimental block of $7.9 \times 10^7$ symbols the projection gives 0.412 Mbit/s, and the blocks of $10^8$ and $10^9$ symbols would raise it to 0.53 and 1.17 Mbit/s, i.e., 36\% and 80\% of the asymptotic rate.}

Table~\ref{tab:comparison} places the present implementation in the context of representative experimental demonstrations of unidimensional CV-QKD. Because the reported experiments employ different channel types, symbol rates, local-oscillator configurations, detection architectures, reconciliation procedures, and security assumptions, the reported SKRs should not be interpreted as a direct performance ranking.

\begin{table*}[t]
    \caption{\label{tab:comparison}
    Comparison with representative experimental demonstrations of
    unidimensional Gaussian-modulated CV-QKD.}
    \begin{ruledtabular}
        \begin{tabular}{lccccc}
            Work & Channel & LO / Detection & Rate
            & Security / Processing & Reported SKR \\
            \hline
            \cite{wang2017experimental}
            & 50 km
            & TLO; homodyne
            & 500 kHz
            & Asym.; $\beta=95.2\%$; RR, PA
            & 0.70 kb/s \\

            \cite{shen2019free}
            & 460 m
            & TLO; homodyne
            & 10 kHz
            & Asym.; $\beta=95\%$ assumed
            & 0.152 kb/s \\

            \cite{nandan2026experimental}
            & Free space
            & TLO; homodyne
            & 2.5 MHz
            & Asym.; $\beta=95\%$ assumed
            & 0.487 Mbit/s \\

            \cite{guo2026pilot}
            & 17.5 km
            & LLO; heterodyne
            & 50 MBaud
            & Asym.; $\beta=97\%$ assumed
            & 6.12 kbit/s \\

            This work
            & 12 km
            & LLO; heterodyne
            & 156.25 MBaud
            & FS; $\beta=93.16\%$; RR, PA
            & \shortstack[c]{
                1.421 Mbit/s (asym.),
                0.364 Mbit/s (FS)
              } \\
        \end{tabular}
    \end{ruledtabular}
    \vspace{2pt}

    \footnotesize
    \noindent
    Asym.: asymptotic; FS: finite-size;
    TLO: transmitted local oscillator;
    LLO: local local oscillator;
    RR: reverse reconciliation;
    PA: privacy amplification.
\end{table*}

Previous experiments established single-quadrature CV-QKD in fiber and free-space links using transmitted-LO, polarization, and more recently LLO-based coherent-detection architectures \cite{gehring2015single,wang2017experimental,shen2019free,nandan2026experimental,guo2026pilot}. The present work extends this progression by combining independent-laser RF-heterodyne reception at 156.25~MBaud with practical MDR/LDPC reconciliation, verification, finite-size parameter estimation, and privacy amplification. At 12~km, the system yields a privacy-amplified rate of $1.421$~Mbit/s under the asymptotic treatment and \textcolor{black}{0.364}~Mbit/s \textcolor{black}{under finite-size treatment.}

Although the present experiment deliberately uses an IQ-modulator-based platform and therefore does not directly demonstrate a reduction in transmitter hardware, single-quadrature state preparation offers a path toward simpler symbol generation, electrical driving, optical modulation, calibration, and stabilization. Future transmitters may therefore exploit architectures based on a single amplitude modulator together with DSP algorithms specifically designed for the 1D signal structure.

\section{Conclusion}
\label{conclusion}

We have experimentally demonstrated a unidimensional Gaussian-modulated CV-QKD system over 12 km of optical fiber, combining independent-laser RF-heterodyne reception with an integrated offline classical post-processing chain. Operating at 156.25 MBaud, the system incorporates practical information reconciliation, parameter estimation, physicality-constrained security analysis, and privacy amplification, yielding a privacy-amplified key rate of 1.421 Mbit/s under the asymptotic treatment and 0.364 Mbit/s when finite-size parameter-estimation uncertainties and privacy-amplification corrections are included.

A central result of this work is the quantitative assessment of finite-size effects in an experimental 1D CV-QKD implementation. The approximately 74\% reduction in key rate highlights the substantial impact of statistical uncertainty on a protocol in which the unmodulated quadrature constrains Eve's information through the covariance matrix's physicality. Despite this penalty, we obtain a positive privacy-amplified key rate using experimentally measured channel parameters and actual reconciliation performance, rather than assumed reconciliation efficiencies. The security analysis is restricted to Gaussian collective attacks and does not constitute a complete composable finite-key security proof.

These results establish an experimental benchmark for integrating and evaluating the performance of unidimensional CV-QKD with independent local oscillators. Although the present implementation does not directly reduce transmitter hardware complexity, it provides a practical foundation for investigating the trade-off between simplified state preparation and the statistical constraints imposed by unidimensional modulation. Future work will focus on reducing excess noise, improving finite-size performance, and implementing transmitter architectures that directly exploit the reduced modulation dimensionality.

\section*{Funding}
This work has been fully funded by the project \textit{Demonstração offline do sistema CV-QKD}, supported by QuIIN - Quantum Industrial Innovation EMBRAPII CIMATEC Competence Center in Quantum Technologies, with financial resources from the PPI IoT/Manufatura 4.0 of the MCTI grant number 053/2023, signed with EMBRAPII. 

\section*{Acknowledgments}
We thank John A. Mora Rodríguez and B. Miranda for valuable discussions.

\section*{Disclosures}
The authors declare that there are no conflicts of interest related to this article.

\section*{Data availability}

The data and source code underlying the results presented in this work may be made available upon formal request, subject to review and approval by the authors.

\bibliography{apssamp}

\appendix

\section{Unidimensional Gaussian-modulated CV-QKD security model}
\label{sec:supp_security}

Usenko and Grosshans \cite{usenko2015unidimensional} introduced the unidimensional Gaussian-modulated CV-QKD protocol and its corresponding security analysis. As described in the manuscript, Alice modulates only one optical-field quadrature, denoted by $X$, while the conjugate quadrature $P$ remains unmodulated. However, the formulation is equivalent under interchange of the two quadratures, i.e., modulating on $P$ while leaving $X$ unmodulated. In the entanglement-based representation, the Alice--Bob covariance matrix can be written as
\begin{equation}
\gamma_{AB}=
\begin{pmatrix}
\sqrt{1+V_M} & 0 & C_X & 0 \\
0 & \sqrt{1+V_M} & 0 & C_P \\
C_X & 0 & V_X^B & 0 \\
0 & C_P & 0 & V_P^B
\end{pmatrix}\text{,}
\label{eq:covariance_matrix}
\end{equation}

\noindent where $V_M$ is Alice's modulation variance, $V_X^B$ and $V_P^B$ are Bob's quadrature variances, and $C_X$ and $C_P$ denote the Alice--Bob correlations in the corresponding quadratures. For the modulated quadrature,
\begin{equation}
    C_X = \sqrt{T_X V_M} (1+V_M)^{1/4}\text{,}
    \label{eq:Cx}
\end{equation}

\noindent and
\begin{equation}
    V_X^B = 1 + T_X \left( V_M+\xi_X \right)\text{,}
    \label{eq:VxB}
\end{equation}

\noindent where $T_X$ is the channel transmittance and $\xi_X$ is the input-referred excess noise associated with the $X$ quadrature.

Since Alice does not modulate the $P$ quadrature, the corresponding Alice--Bob correlation $C_P$ cannot be directly estimated from Alice's modulation data. Instead, Bob's measured variance $V_P^B$, together with the covariance matrix's physicality condition, constrains the admissible values of $C_P$. A physical covariance matrix must satisfy
\begin{equation}
    \gamma_{AB}+i\Omega \geq 0\text{,}
    \label{eq:physicality}
\end{equation}

\noindent where $\Omega$ is the symplectic form. Equivalently, all symplectic eigenvalues of $\gamma_{AB}$ must be greater than or equal to unity. For the unidimensional protocol, the physicality constraint can be expressed as
\begin{equation}
    \left( C_P-C_0 \right)^2 \leq \frac{V_M}{\sqrt{1+V_M}} \left( 1-T_X V_0^B \right) \left( V_P^B-V_0^B \right)\text{,}
    \label{eq:physicality_parabola}
\end{equation}
\noindent with
\begin{equation}
    V_0^B = \frac{1}{1+T_X\xi_X}\text{,}
    \label{eq:V0}
\end{equation}
\noindent and
\begin{equation}
    C_0 = - V_0^B \frac{\sqrt{T_XV_M}}{(1+V_M)^{1/4}}\text{.}
    \label{eq:C0}
\end{equation}

The physicality diagram presented in the main manuscript illustrates the admissible region in the $(C_P,V_P^B)$ plane defined by \eqref{eq:physicality_parabola}. We obtain its boundary by replacing the inequality with an equality, with $V_0^B$ and $C_0$ determined by Eqs.~\eqref{eq:V0} and~\eqref{eq:C0}, respectively. Accordingly, the figure in the main manuscript graphically represents the physicality constraint derived in this section, rather than an additional independent condition.

The experimentally determined quantities therefore define an admissible interval of $C_P$. Eve's Holevo information is conservatively evaluated by maximizing over all physically allowed correlations,
\begin{equation}
    \chi_{BE}^{\mathrm{worst}} = \max_{C_P \in \mathcal{P}} \chi_{BE}\text{,}
    \label{eq:holevo_optimization}
\end{equation}

\noindent where $\mathcal{P}$ denotes the physical region defined by Eq.~\eqref{eq:physicality_parabola}. In the finite-size analysis (Sec.~\ref{sec:supp_pe}), the confidence intervals for $T_X$, $\xi_{B,X}$, and $V_P^B$ are also considered. The optimization then extends over the union of physical regions consistent with these statistically allowed parameter values, as conceptually indicated in the finite-size treatment of the physicality diagram in the main manuscript.


\textcolor{black}{\section{Holevo bound with a private P quadrature}}

\textcolor{black}{Bob's RF-heterodyne receiver recovers both quadratures of every symbol. Only X-quadrature outcomes $x_{B}$ form the raw key: they are the input of reverse reconciliation and privacy amplification. The P-quadrature outcomes $p_B$ are used solely to estimate $V_P^B$ in \eqref{eq:covariance_matrix} and are never disclosed, so they do not add to the information accessible to Eve about the key. Under Gaussian collective attacks, Eve's information on the key is therefore bounded by the Holevo quantity
\begin{equation}
    \chi_{BE} = \chi(x_B:E) = S(\rho_E) - S(\rho_{E|x_B}),
\end{equation}
\noindent where $S(\rho)$ denotes the von Neumann entropy \cite{laudenbach2018continuous}. This criterion differs from the heterodyne-based unidimensional protocol of Ref.~\cite{usenko2025shared}, in which both heterodyne outcomes are processed by Bob and Eve's information is bounded by the larger quantity $\chi(x_B,p_B:E) \geq \chi(x_B:E)$.}

\begin{figure*}[t]
    \centering
    \includegraphics[width=\textwidth]{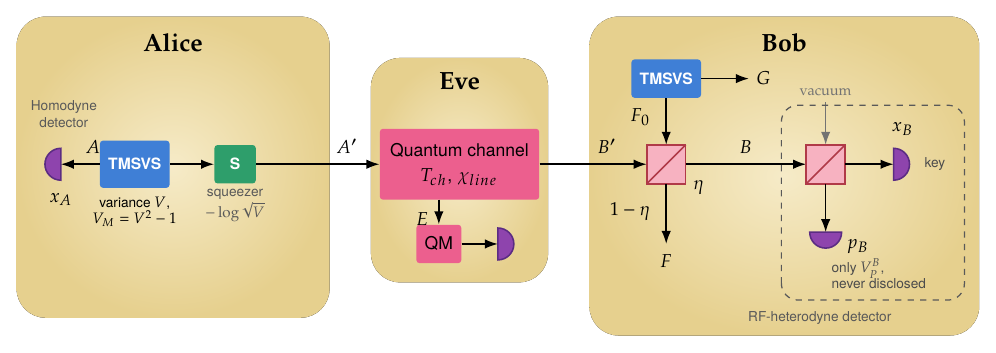}
    \caption{\justifying \textcolor{black}{Trusted-noise model for the unidimensional coherent-state CV-QKD protocol in the entanglement-based representation~\cite{usenko2015unidimensional}. Alice and Bob share a two-mode squeezed vacuum state (TMSVS) of variance $V$, with $V_M = V^2 - 1$. Alice performs a homodyne detection of $x_A$ on her mode $A$, which conditionally prepares coherent states displaced along the $x$ quadrature only; the other mode is squeezed by $S$ (squeezing parameter $-\log\sqrt{V}$) and sent to Bob through the quantum channel as mode $A'$. Eve interacts with this mode, leading to an output mode $B'$, and keeps her results in a quantum memory (QM). At Bob's station, $B'$ interacts with one mode, $F_0$, of another TMSVS by mixing them in a beam splitter of transmittance $\eta$, which models the detection efficiency and the electronic noise; the mode $F$ is discarded and the mode $G$ remains with Bob. The resulting mode $B$ is measured by RF-heterodyne detection. Only the $x_B$ outcomes form the raw key; the $p_B$ outcomes are used solely to estimate $V_P^B$ and are never disclosed.}}
    \label{fig:trustednoise}
\end{figure*}

\textcolor{black}{We also consider the trusted-noise scenario, where the detection efficiency $\eta_B$ and the electronic noise $\nu_{el}$ are treated as trusted \cite{wang2017finite,anka2026introductory} (see Fig.~\ref{fig:trustednoise} for a schematic representation). Unlike Ref.~\cite{usenko2025shared}, Bob only disclose the $x_B$ outcome. Each heterodyne detection carries half of the received signal, so the X marginal of the heterodyne measurement is equivalent to a homodyne measurement of efficiency $\eta_B/2$ with electronic noise $\nu_{el}$, where $\nu_{el}$ is referred to the shot noise of one output (per quadrature). We model it by a beam-splitter of transmittance $\eta_B/2$ that mixes mode $B$ with one mode $F_0$ of a two-mode squeezed vacuum $F_0G$ of quadrature variance $\nu = 1 + \nu_{el}/(1 - \eta_B/2)$, followed by a homodyne measurement of $x$ on the transmitted mode, whose variance is $(\eta_B/2)V_X^B + 1 - \eta_B/2 + \nu_{el}$. The modes $F_0G$ purify the system, so that $S(\rho_E) = S(\rho_{AB})$ and $S(\rho_{E|x_B}) = S(\rho_{AF_0G|x_B})$. The von Neumann entropy follow from sympletectic eigenvalues, $S(\rho) = \sum_i g[(\nu_i - 1)/2]$ with $g(x) = (x + 1) \log_2(x + 1) - x log_2x$. The symplectic eigenvalues $\nu_{1,2}$ and $\nu_{3,4,5}$ are computed from $\gamma_{AB}$ and $\gamma_{AFG|x_B}$, respectively, as \cite{wang2017finite}
\begin{equation}
    \nu^2_{1,2} = \dfrac{A \pm \sqrt{A^2 - 4B}}{2},~~~\nu^2_{3,4} = \dfrac{C \pm \sqrt{C^2 - 4D}}{2}, ~~~\text{and}~~~\nu_5 = 1,
\end{equation}
where
\begin{align}
A &= \det\gamma_A + \det\gamma_B + 2\det C \nonumber\\
  &= 1 + V_M + V_P^B \nonumber\\
  &\quad + V_P^B\,T_X(V_M + \xi_X) \nonumber\\
  &\quad + 2C_P\sqrt{T_X V_M}\,(1+V_M)^{1/4},
  \label{eq:A}\\
B &= \det\gamma_{AB} \nonumber\\
  &= (1 + T_X\xi_X)\left[(1+V_M)\,V_P^B \right. \nonumber\\
  &\quad \left. - C_P^2\sqrt{1+V_M}\right],
  \label{eq:B}
\end{align}
and
\begin{align}
C &= \frac{A\,(1 + v_{\mathrm{el}}) + \left[(1 + T_X\xi_X)(V_M + 2) + V_M T_X - A\right]\eta_B/2}{d_X}, \label{eq:C}\\
D &= \frac{B\,(1 + v_{\mathrm{el}} - \eta_B/2) + (1+V_M)(1 + T_X\xi_X)\,\eta_B/2}{d_X}, \label{eq:D}
\end{align}
with $d_X = 1 + v_{el} + (\eta_B/2) T_X (V_M + \xi_X)$ the variance of Bob's X outcome. The Holevo bound is then
\begin{equation}
\chi_{BE} = g\!\left[\tfrac{\nu_1 - 1}{2}\right] + g\!\left[\tfrac{\nu_2 - 1}{2}\right]
          - g\!\left[\tfrac{\nu_3 - 1}{2}\right] - g\!\left[\tfrac{\nu_4 - 1}{2}\right].
\label{eq:chiclosed}
\end{equation}}

\textcolor{black}{Finally, the secret key fraction is given by
\begin{equation}
K = \beta I_{AB} - \chi_{BE},
\end{equation}
where the mutual information can be written as $I_{AB} = \dfrac{1}{2}\log_2(1 + \text{SNR})$, with
\begin{equation}
\text{SNR} = \dfrac{\eta_B T_X V_M/2}{1 + \nu_{el} + \eta_B T_X \xi_X/2}.
\end{equation}}


\section{Parameter estimation and finite-size statistical bounds}
\label{sec:supp_pe}

Parameter estimation uses the full set of recovered quantum symbols submitted to information reconciliation, including symbols from LDPC blocks that do not decode successfully \cite{jain2022practical}. Consequently, decoding failures contribute to the frame-error rate but do not reduce the number of samples used for parameter estimation. For the modulated quadrature, the linear channel model is written as \cite{leverrier2010finite}
\begin{equation}
    x_B=t_X x_A+z_X\text{,}
    \label{eq:linear_channel}
\end{equation}

\noindent where $x_A$ denotes Alice's Gaussian modulation values, $x_B$ denotes Bob's recovered measurements, $t_X$ is the effective channel gain, and $z_X$ represents the total additive noise.

The channel gain, channel transmittance, and residual-noise variance are estimated as
\begin{equation}
\begin{split}
\hat{t}_X &=
\frac{\operatorname{Cov}(x_A,x_B)}
{\operatorname{Var}(x_A)},\\
\hat{T}_X &= \frac{2\hat{t}_X^2}{\eta_B},\\
\hat{\sigma}_X^2 &=
\operatorname{Var}\left(x_B-\hat{t}_Xx_A\right),
\end{split}
\label{eq:channel_estimators}
\end{equation}

\noindent where the factor of two accounts for the RF heterodyne detection convention and $\eta_B=0.80$ is Bob's detection efficiency.

We estimate the Bob-referred excess noise after subtracting the calibrated shot-noise and electronic-noise contributions. For RF heterodyne detection, the Bob-referred excess noise is related to the input-referred excess noise used in the covariance-matrix model by
\begin{equation}
    \xi_{B,X}=\frac{\eta_BT_X}{2}\xi_X\text{.}
    \label{eq:bob_referred_noise}
\end{equation}

\noindent \textcolor{black}{Each heterodyne output receives half of the received signal and half of the excess noise, with the shot noise of one output normalized to unity \cite{pietri2024qosst}.} The main manuscript reports the input-referred quantity $\xi_X$.

For the final experimental data set, the parameter-estimation block contains $N_{\mathrm{PE}}=78{,}562{,}560$ recovered quantum symbols. The corresponding point estimates for the modulated quadrature are
\begin{equation}
\begin{split}
T_X &\simeq 0.5740,\\
\xi_X &= 31.456~\mathrm{mSNU},\\
\xi_{B,X} &\simeq 7.222~\mathrm{mSNU},\\
V_P^B &= 1.01283~\mathrm{SNU}.
\end{split}
\label{eq:experimental_asym}
\end{equation}

\noindent The Bob-referred and input-referred noise values are consistent with Eq.~\eqref{eq:bob_referred_noise}. The unmodulated-quadrature variance $V_P^B$ is estimated separately from Bob's recovered $P$-quadrature measurements and enters the physicality-constrained optimization described in Sec.~\ref{sec:supp_security}.

In the finite-size regime, the statistical analysis constructs confidence intervals for the channel transmittance, excess noise, and unmodulated-quadrature variance. RF heterodyne detection provides both $X$ and $P$ measurements for each recovered symbol. Accordingly, the numbers of observations used in the two quadrature branches are $m=l=N_{\mathrm{PE}}=78{,}562{,}560$; these are measurements of the same symbol set, not disjoint subsets.

Following the adopted finite-size parameter-estimation treatment, we set
\begin{equation}
    \delta_{\mathrm{PE}}^{\mathrm{1D}}=10^{-10}\text{,} \qquad
    z_{\mathrm{PE}}=\sqrt{2}\,\operatorname{erfc}^{-1}\left(\delta_{\mathrm{PE}}\right)\text{.}
    \label{eq:pe_security_parameters}
\end{equation}

The resulting finite-size confidence region can be written as
\begin{equation}
\begin{split}
    T_X^{\min} &\leq T_X\leq T_X^{\max}\text{,}\\
    \xi_{B,X} &\leq \xi_{B,X}^{\max}\text{,}\\
    V_{P,\min}^B &\leq V_P^B\leq V_{P,\max}^B\text{.}
\end{split}
\label{eq:finite_confidence_region}
\end{equation}
\textcolor{black}{With $\hat t_X = \sqrt{\eta_B\hat T_X/2}$, $\hat\sigma_X^2 = 1 + v_{\mathrm{el}} + \eta_B\hat T_X\hat\xi_X/2$ and $\hat v_P = 1 + v_{\mathrm{el}} + (\eta_B/2)(V_P^B - 1)$ the Bob-referred $P$ variance, the bounds are
\begin{equation}
\begin{split}
T_X^{\min,\max}
&= \frac{2}{\eta_B}
\left(\hat t_X \mp z_{\mathrm{PE}}
\sqrt{\frac{\hat\sigma_X^2}{m V_M}}\right)^{\!2},\\
\xi_{B,X}^{\max}
&= \hat\sigma_X^2
\left(1 + z_{\mathrm{PE}}\sqrt{\frac{2}{m}}\right)\\
&\quad - 1 - v_{\mathrm{el}},\\
V_{P,\min/\max}^B
&= 1 + \frac{2}{\eta_B}
\left[\hat v_P
\left(1 \mp z_{\mathrm{PE}}\sqrt{\frac{2}{m}}\right)\right.\\
&\qquad \left. - 1 - v_{\mathrm{el}}\right].
\end{split}
\label{eq:bounds}
\end{equation}
which give $T_X \in [0.5724, 0.5756]$, $\xi_{B,X}^{\max} = 8.297$~mSNU and $V_P^B \in [1.01015, 1.01552]$ for the final data set.
}

For a given transmittance within the confidence interval, the corresponding input-referred excess-noise upper bound is
\begin{equation}
    \xi_X^{\max}(T_X)=\frac{2\xi_{B,X}^{\max}}{\eta_BT_X}\text{.}
    \label{eq:finite_input_noise}
\end{equation}

Rather than assuming, at first glance, that either boundary of the transmittance confidence interval is the worst-case channel, Eve's Holevo information is maximized over the finite-size confidence region, together with the physically admissible Alice--Bob correlation $C_P$ in the unmodulated quadrature. Defining $\mathcal{C}_T$, $\mathcal{C}_{\xi}$, and $\mathcal{C}_P$ as the statistically admissible sets for $T_X$, $\xi_{B,X}$, and $V_P^B$, respectively, with $\mathcal{C}_{\xi}$ restricted by the upper confidence bound $\xi_{B,X}^{\max}$, we obtain
\begin{equation}
    \chi_{BE}^{\mathrm{finite}} =\max_{\substack{ T_X\in\mathcal{C}_T,\;\xi_{B,X}\in\mathcal{C}_{\xi},\\ V_P^B\in\mathcal{C}_P,\\ C_P\in\mathcal{P}(T_X,\xi_X,V_P^B) }}\chi_{BE}\text{,} \label{eq:finite_holevo}
\end{equation}

\noindent where $\xi_X=2\xi_{B,X}/(\eta_BT_X)$ at each candidate operating point and $\mathcal{P}$ is the physical region defined by \eqref{eq:physicality_parabola}. The worst-case values of $T_X$, $\xi_X$, and $V_P^B$, as well as $\chi_{BE}^{\mathrm{finite}}$, are determined jointly by this constrained optimization. \textcolor{black}{For the final data set, the maximum is attained at the upper transmittance bound, $T_X = 0.5756$, with $\xi_X = 36.04$,SNU from \eqref{eq:finite_input_noise}, $V_P^B = 1,01552$ and $C_P = -0.5694$: both the noise bound of \eqref{eq:finite_input_noise} and the region defined by \eqref{eq:physicality_parabola} depend on $T_X$, and the larger admissible correlation at the upper bound overweighs the smaller channel loss.}

We obtain the reconciliation contribution from the measured information-reconciliation performance, rather than reevaluating it at an extremal point of the parameter-estimation confidence region. For the final data set, the reconciliation efficiency is $\beta\simeq0.9316$, and we use $\beta I_{AB}$ consistently in the asymptotic and finite-size key-rate calculations. In the finite-size parameter-estimation comparison considered here, the statistical confidence region modifies the worst-case Holevo information.

\textcolor{black}{Finally, we present the effects of the privacy-amplification correction. The finite length of the key also reduces the number of secret bits that can be extracted by privacy amplification. Following Leverrier \textit{et al.}~\cite{leverrier2010finite}, the finite-size secret fraction is
\begin{equation}
\begin{aligned}
K_{\mathrm{finite}} &= \beta I_{AB} - \chi_{BE}^{\mathrm{finite}} - \Delta(n),\\
\Delta(n) &= (2\dim\mathcal{H}_X + 3)\sqrt{\frac{\log_2(2/\bar\varepsilon)}{n}} = 7\sqrt{\frac{\log_2(2/\bar\varepsilon)}{n}},
\end{aligned}
\label{eq:deltan}
\end{equation}
where $\dim\mathcal{H}_X = 2$ is the dimension of the binary key variable after reconciliation, $\bar\varepsilon$ is the smoothing parameter of the min-entropy and $n$ is the number of symbols that enter privacy amplification. This is the form adopted in the finite-size analysis of the unidimensional protocol of Ref.~\cite{wang2017finite}; the additional term $(2/n)\log_2(1/\varepsilon_{\mathrm{PA}})$ of Ref.~\cite{leverrier2010finite}, which accounts for the failure probability $\varepsilon_{\mathrm{PA}}$ of privacy amplification, equals $1.3\times10^{-6}$~bit/symbol for the present block with $\varepsilon_{\mathrm{PA}} = 10^{-10}$ and is omitted. Since only the successfully reconciled blocks are hashed, $n = (1 - \mathrm{FER})\,N_{\mathrm{key}} = 52{,}933{,}920$ symbols (442 accepted LDPC blocks of 119,760 symbols), and with $\bar\varepsilon = 10^{-10}$,
\begin{equation}
\Delta(n) = 5.628\times10^{-3}~\mathrm{bit/symbol}.
\label{eq:deltavalue}
\end{equation}
The finite-size secret fraction of the final data set is therefore $K_{\mathrm{finite}} = 0.16699 - 0.15687 - 0.00563 = 0.00449$~bit/symbol, against $K = 0.01602$~bit/symbol in the asymptotic treatment: the parameter-estimation confidence region and $\Delta(n)$ remove $5.90\times10^{-3}$ and $5.63\times10^{-3}$~bit/symbol, respectively. The treatment bounds collective Gaussian attacks with the parameter-estimation and privacy-amplification corrections of Ref.~\cite{leverrier2010finite}; it is not a composable finite-key proof.}

\end{document}